\documentclass[english,twocolumn,prl,superscriptaddress]{revtex4-1}
\usepackage[T1]{fontenc}
\usepackage[latin9]{inputenc}
\usepackage{color}
\usepackage{amsmath}
\usepackage{amssymb}
\usepackage{graphicx}
\usepackage{esint}
\PassOptionsToPackage{normalem}{ulem}
\usepackage{ulem}

\makeatletter
\usepackage{hyperref}
\usepackage{braket}

\newcommand{\noise}[1]{\mathrel{\langle\mskip-3.5mu\langle #1 \rangle\mskip-3.5mu\rangle}}

\makeatother

\usepackage{babel}
\begin{document}

\title{Stability of dynamical topology against dynamical noise in quantum quenches}

\author{Lin Zhang}
\affiliation{International Center for Quantum Materials and School of Physics, Peking University, Beijing 100871, China}
\affiliation{Collaborative Innovation Center of Quantum Matter, Beijing 100871, China}
\author{Long Zhang}
\affiliation{International Center for Quantum Materials and School of Physics, Peking University, Beijing 100871, China}
\affiliation{Collaborative Innovation Center of Quantum Matter, Beijing 100871, China}
\author{Xiong-Jun Liu}
\thanks{Correspondence addressed to: xiongjunliu@pku.edu.cn}
\affiliation{International Center for Quantum Materials and School of Physics, Peking University, Beijing 100871, China}
\affiliation{Collaborative Innovation Center of Quantum Matter, Beijing 100871, China}
\affiliation{Beijing Academy of Quantum Information Science, Beijing 100193, China}
\affiliation{CAS Center for Excellence in Topological Quantum Computation, University of Chinese Academy of Sciences, Beijing 100190, China}
\affiliation{Institute for Quantum Science and Engineering and Department of Physics, Southern University of Science and Technology, Shenzhen 518055, China}

\begin{abstract}
Equilibrium topological phases are robust against weak static disorder but may break down in the strong disorder regime. Here we explore the stability of the quench-induced emergent dynamical topology in the presence of dynamical noise. We develop an analytic theory and show that for weak noise, the quantum dynamics induced by quenching an initial trivial phase to Chern insulating regime exhibits robust emergent topology on certain momentum subspaces called band inversion surfaces (BISs). The dynamical topology is protected by the minimal oscillation frequency over the BISs, mimicking a bulk gap of the dynamical phase. Singularities emerge in the quench dynamics, with the minimal oscillation frequency vanishing on the BISs if increasing noise to critical strength, manifesting a dynamical topological transition, beyond which the emergent topology breaks down. Two types of dynamical transitions are predicted. Interestingly, we predict a {\em sweet spot} in the critical transition when noise couples to all three spin components in the same strength, in which case the dynamical topology survives at arbitrarily strong noise regime. This work unveils novel features of the dynamical topology under dynamical noise, which can be probed with control in experiment.
\end{abstract}
\maketitle

\emph{Introduction}.---Over the last decades, the topological quantum phases which host topologically protected boundary modes through the celebrated bulk-boundary correspondence have ignited extensive research in condensed matter physics~\cite{Hasan2010,Qi2011,Yan2012,Chiu2016,Yan2017},
photonic systems~\cite{Haldane2008,Rechtsman2013,Khanikaev2013,Ozawa2019}
and ultracold atoms~\cite{Aidelsburger2013,Miyake2013,Liu2013,Liu2014,Jotzu2014,Aidelsburger2015,Wu2016}.
With the full controllability, the quantum simulation of topological phases facilitates
the inverstigation of topological physics out of equilibrium~\cite{Caio2015,Hu2016,Wang2017,Flaschner2018,Song2018,McGinley2019,Hu2020,CaiHan2019,Lu2019,QiuX2019,XiePRL2020,Yu2020}.
In particular, the dynamical bulk-surface correspondence was proposed~\cite{Zhang2018,Zhang2019,Zhang2019b,Zhang2019c,XLYu2020},
showing that the bulk topology of an equilibrium $d$-dimensional ($d$D) topological
phase universally corresponds to the dynamical topological patterns, emerging in quantum dynamics induced by quench,
on the $(d-1)$D momentum subspaces called band-inversion surfaces (BISs). The dynamical bulk-surface
correspondence provides new schemes to characterize and detect the equilibrium topology via non-equilibrium quench dynamics, with
the experimental verifications having been widely reported recently~\cite{Sun2018b,Yi2019,Wang2019,Song2019,Ji2020,Xin2020,Niu2020}.

The emergent dynamical topology on BISs is primarily studied in the unitary quench dynamics
without considering the inevitable dynamical noise in practical experiments~\cite{Yi2019,Wang2019,Sun2018b,Song2019,Ji2020,Xin2020,Niu2020},
which even drives the quantum dynamics into dissipative regime~\cite{Carmichael1993}.
It is widely studied that the equilibrium topological phases are robust against weak static disorder
but are destroyed by strong disorder with bulk gap closed~\cite{Sheng2006,Li2009,Prodan2010}.
As an analogy to static disorder effect, is the emergent dynamical topology in quantum quenches
robust against dynamical noise? If so, what is the protection mechanism?
Answering these questions shall promote the basic understanding of the emergent topology
in quench dynamics and stimulate the experimental studies of its stability with controllable noise.

In this work, we address these issues by investigating the quench-induced dynamical topology under dynamical noise
for 2D quantum anomalous Hall (QAH) systems. The analytic theory of the quench dynamics is developed based on the stochastic Schr\"{o}dinger equation. In the weak noise regime, we show that the dynamical topology emerging
on the dynamical band-inversion surfaces (dBISs) is protected by the minimal oscillation frequency which mimics the bulk gap of the emergent dynamical phase. In the strong noise regime, the singularities emerge in the dynamical patterns on dBISs and the emergent topology breaks down. Two types of critical points of the dynamical transition are obtained, with a {\em sweet spot} regime being uncovered, in which the dynamical topology survives at arbitrary strong noise.

\begin{figure}[t]
\includegraphics{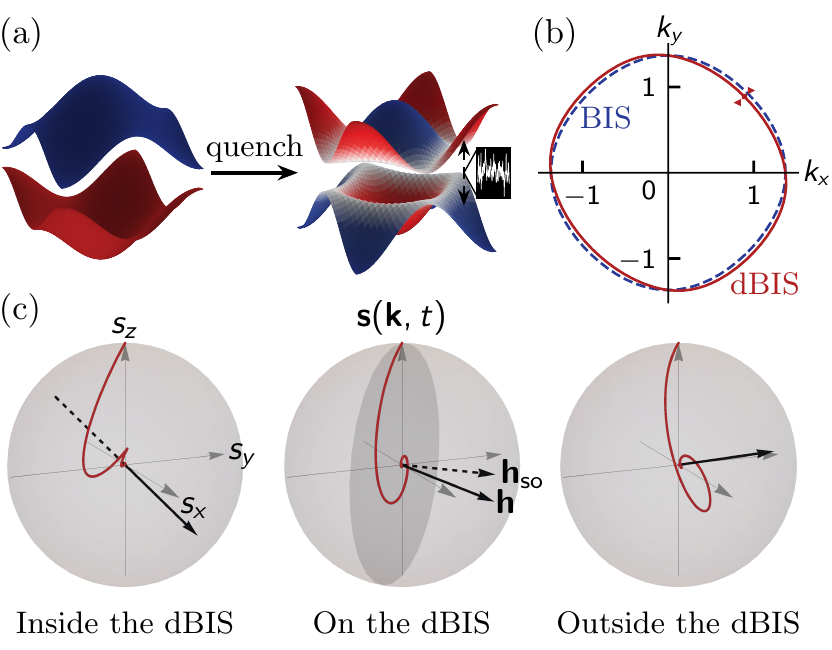}
\caption{Noise-induced dissipative quench dynamics and the dynamical band-inversion
surfaces. (a) Quenching a 2D QAH model from deep
trivial state to the topologically nontrivial regime. The white noise
(the insert) randomly couples the upper and lower bands.
(b) The dBIS (red line) and BIS (dashed blue line). (c) The stochastic averaged
spin polarizations $\boldsymbol{s}(\mathbf{k},t)$ for the momenta marked
in (b) by red triangles. On the dBIS, the evolution of spin polarization
(red line) is always within the plane perpendicular to the SO axis $\mathbf{h}_{{\rm so}}$
(black dashed arrow), while it spirals around the post-quench Hamiltonian vector $\mathbf{h}$ (black arrows)
for momenta off the dBIS (including the BIS). Here we set $m_{z}=1.2t_{0}$ and $t_{{\rm so}}=0.2t_{0}$
for the post-quench Hamiltonian. The noise strength is $w_{x}=0.12t_{0}$,
$w_{y}=0$, $w_{z}=0.02t_{0}$.\label{fig:figure1}}
\end{figure}

\emph{The model}.---We start with 2D QAH models coupling to dynamical white noise, with the Hamiltonian
\begin{equation}
\tilde{H}(\mathbf{k},t)=H(\mathbf{k})+\boldsymbol{w}(\mathbf{k},t)\cdot\boldsymbol{\sigma}+\delta m_{z}(t)\sigma_{z},
\end{equation}
where $H(\mathbf{k})=\mathbf{h}(\mathbf{k})\cdot\boldsymbol{\sigma}$ gives QAH phase,
and $w_{i}(\mathbf{k},t)$ is magnetic white noise of strength $\sqrt{w_{i}}$, which exists and is controllable in experiments~\cite{Sun2018b,Yi2019,Wang2019,Song2019,Ji2020}, satisfying
$\noise{w_{i}(\mathbf{k},t)}=0$ and $\noise{w_{i}(\mathbf{k},t)w_{j}(\mathbf{k},t')}=w_{i}\delta_{ij}\delta(t-t')$.
Here $\noise{\cdot}$ denotes the stochastic average over different noise configurations.
We tune the magnetization $\delta m_z$ to trigger the quench dynamics from the deep trivial
state to the topologically nontrivial phase, see Fig.~\ref{fig:figure1}(a). For $t<0$, we set $\delta m_z\ll 0$
such that the system is initially prepared in the fully polarized state
$\ket{\psi(\mathbf{k},t=0)}=\ket{\uparrow}$ and the noise is negligible for the pre-quench state.
Quenching $\delta m_{z}\to0$ at $t=0$ leads to evolution under the post-quench Hamiltonian with noise
$H_{\rm post}(\mathbf{k},t)=H(\mathbf{k})+\boldsymbol{w}(\mathbf{k},t)\cdot\boldsymbol{\sigma}$.
The dynamics is governed by the stochastic Schr\"odinger equation~\cite{Diosi1988,Barchielli1991,Barchielli2009,Semina2014},
$\mathrm{i}\partial_{t}\ket{\psi(\mathbf{k},t)}=H_{\rm post}(\mathbf{k},t)\ket{\psi(\mathbf{k},t)}$,
a noise-induced random unitary evolution.
Our study can be readily generalized to quench dynamics along an arbitrary axis~\cite{Zhang2018}
and from a generic trivial state~\cite{Zhang2019b}.

The quench dynamics with noise can be quantified by the stochastic averaged spin polarization
\begin{equation}
\boldsymbol{s}(\mathbf{k},t)\equiv\noise{\braket{\psi(\mathbf{k},t)\vert\boldsymbol{\sigma}\vert\psi(\mathbf{k},t)}}.
\end{equation}
The random evolution of spin polarization renders the dissipative quantum dynamics [Fig.~\ref{fig:figure1}(c)],
which can be obtained from the Lindblad master equation~\cite{Barchielli2009,Semina2014,Lindblad1976} for the stochastic averaged
density matrix $\boldsymbol{s}(\mathbf{k},t)=\mathrm{Tr}[\rho(\mathbf{k},t)\boldsymbol{\sigma}]$, with $\rho(\mathbf{k},t)=\noise{\ket{\psi(\mathbf{k},t)}\bra{\psi(\mathbf{k},t)}}$ and
\begin{equation}
\partial_{t}\rho(\mathbf{k},t) = \mathcal{L_{\mathbf{k}}}[\rho(\mathbf{k},t)],\ \partial_{t}{\boldsymbol{s}}(\mathbf{k},t)=\mathcal{L}(\mathbf{k})\boldsymbol{s}(\mathbf{k},t).\label{eq:master equation}
\end{equation}
Here the Liouvillian superoperator reads~\cite{SM} 
\begin{equation}
\mathcal{L}(\mathbf{k})=2\begin{bmatrix}-w_{y}-w_{z} & -h_{z}(\mathbf{k}) & h_{y}(\mathbf{k})\\
h_{z}(\mathbf{k}) & -w_{x}-w_{z} & -h_{x}(\mathbf{k})\\
-h_{y}(\mathbf{k}) & h_{x}(\mathbf{k}) & -w_{x}-w_{y}
\end{bmatrix},\label{eq:Liouvillian superoperator}\nonumber
\end{equation}
where the noise $w_j$ is directly coupled to spin components $s_{i}$ ($i\neq j$) for its dephasing effect in the basis $\ket{\sigma_j}$.
The noise in general couples the eigenstates of $H(\mathbf{k})$, leading to a decay effect
instead of pure dephasing and modulating the oscillation frequency.
The same master equation can be reached for quantum noise~\cite{Gardiner2004}, so this study is applicable to the case with quantum noise.

The solution to the dissipative quench dynamics can be generically written as~\cite{SM}
$\boldsymbol{s}(\mathbf{k},t)=\boldsymbol{s}_{0}(\mathbf{k})e^{-\lambda_{0}(\mathbf{k})t}++\boldsymbol{s}_{+}(\mathbf{k})e^{-[\lambda_{1}(\mathbf{k}) +\mathrm{i}\omega(\mathbf{k})]t}+\boldsymbol{s}_{-}(\mathbf{k})e^{-[\lambda_{1}(\mathbf{k})-\mathrm{i}\omega(\mathbf{k})]t}$,
with the coefficients  $\boldsymbol{s}_{\alpha}(\mathbf{k})=[\boldsymbol{s}_{\alpha}^{L}(\mathbf{k})\cdot\boldsymbol{s}(\mathbf{k},0)]\boldsymbol{s}_{\alpha}^{R}(\mathbf{k})$
for $\alpha=0,\pm$. Here $\boldsymbol{s}_{\alpha}^{L(R)}$ satisfying $\boldsymbol{s}_{\alpha}^{L}(\mathbf{k})\cdot\boldsymbol{s}_{\beta}^{R}(\mathbf{k})=\delta_{\alpha\beta}$
are the left (right) eigenvectors of the Liouvillian superoperator
\begin{eqnarray}
\mathcal{L}^{T}(\mathbf{k})\boldsymbol{s}^{L}_{\alpha}=-\lambda_{\alpha}\boldsymbol{s}^{L}_{\alpha}, \
\mathcal{L}(\mathbf{k})\boldsymbol{s}^{R}_{\alpha}=-\lambda_{\alpha}\boldsymbol{s}^{R}_{\alpha},
\end{eqnarray}
with eigenvalues $\lambda_{0}$ and $\lambda_{\pm}=\lambda_{1}\pm\mathrm{i}\omega$ respectively, and $\boldsymbol{s}^{L(R)}_0$ are real.
The coefficients $\lambda_{0,1}$ reflect explicitly the dissipation of the spin dynamics due to the dynamical noise. As is known that without noise, the nontrivial topology emerges in the quench dynamics on BISs where the full spin-flip oscillations occur and thus the time-averaged spin-polarization $\overline{\boldsymbol{s}(\mathbf{k})}\equiv\lim_{T\to\infty}\frac{1}{T}\int^{T}_{0} \mathrm{d}t\,\boldsymbol{s}(\mathbf{k},t)$ vanishes~\cite{Zhang2018,Zhang2019,Zhang2019b}. This characterization needs generalization for the noise regime, in which the decay dominates the long-time evolution and the direct time-averaging $\overline{\boldsymbol{s}(\mathbf{k})}$ vanishes at any $\bold k$ and does not capture the topology. Below we first extend the characterization theory to the generic case with dissipation.

\emph{Emergent dynamical topology}.---For unitary quench dynamics, the BISs correspond to the momentum subspace
$\{\mathbf{k}\vert h_{z}(\mathbf{k})=0\}$ with $\mathbf{h}(\mathbf{k})\cdot\boldsymbol{s}(\mathbf{k},0)=0$, where the spin processes perpendicularly to spin-orbit (SO) axis $\mathbf{h}_{{\rm so}}(\mathbf{k})=(h_x,h_y)$.
With dynamical noise, on such BISs the spin dynamics are complicated. However, we show that
on the $\bold k$-subspace dubbed dBISs defined by $\boldsymbol{s}_{0}^{L}$ as
\begin{equation}
{\rm dBIS}\equiv\{\mathbf{k}\vert\boldsymbol{s}_{0}^{L}(\mathbf{k})\cdot\boldsymbol{s}(\mathbf{k},0)=0\},\label{eq:dBIS}
\end{equation}
under noise the spin processes fully in the plane perpendicular to the SO axis $\mathbf{h}_{\rm so}$~\cite{SM}. 
Away from the dBISs, the spin evolves along a generic 3D curve in the Bloch sphere.
For the initial state $\ket{\uparrow}$, the dBISs are analytically given by the momenta satisfying
$h_{z}=(w_{y}-w_{x})h_{x}h_{y}/(h_{x}^{2}+h_{y}^{2})$ with $\boldsymbol{s}^L_0\sim(h_x,h_y,0)$.
The dBISs return to the BISs if the noise strengths are zero or $w_x=w_y$.
The numerical results in Fig.~\ref{fig:figure1} confirm features on dBIS, where we consider the QAH model~\cite{Liu2014,Baozong} realized in recent experiments~\cite{Wu2016,Sun2018},
$H(\mathbf{k})=\mathbf{h}(\mathbf{k})\cdot\boldsymbol{\sigma}$
with $\mathbf{h}(\mathbf{k})=(t_{{\rm so}}\sin k_{x},t_{{\rm so}}\sin k_{y},m_{z}-t_{0}\cos k_{x}-t_{0}\cos k_{y})$,
where $t_{0}$ and $t_{{\rm so}}$ denote the spin-conserved
and spin-flipped hopping. The magnetization $m_{z}$ is suddenly changed from a large negative value to $m_{z}=1.2t_{0}$ [Fig.~\ref{fig:figure1}(a)], together with $t_{{\rm so}}=0.2t_{0}, w_{x}=0.12t_{0}$, $w_{y}=0$, and $w_{z}=0.02t_{0}$. The dBIS differs from the BIS even the noise is weak [Fig.~\ref{fig:figure1}(b)]. 
On the dBIS, the spin polarization evolves exactly within
the plane perpendicular to $\mathbf{h}_{{\rm so}}$, but 
not for momenta off the dBIS [including the BIS, see Fig.~\ref{fig:figure1}(c)].

\begin{figure*}[ht]
\includegraphics{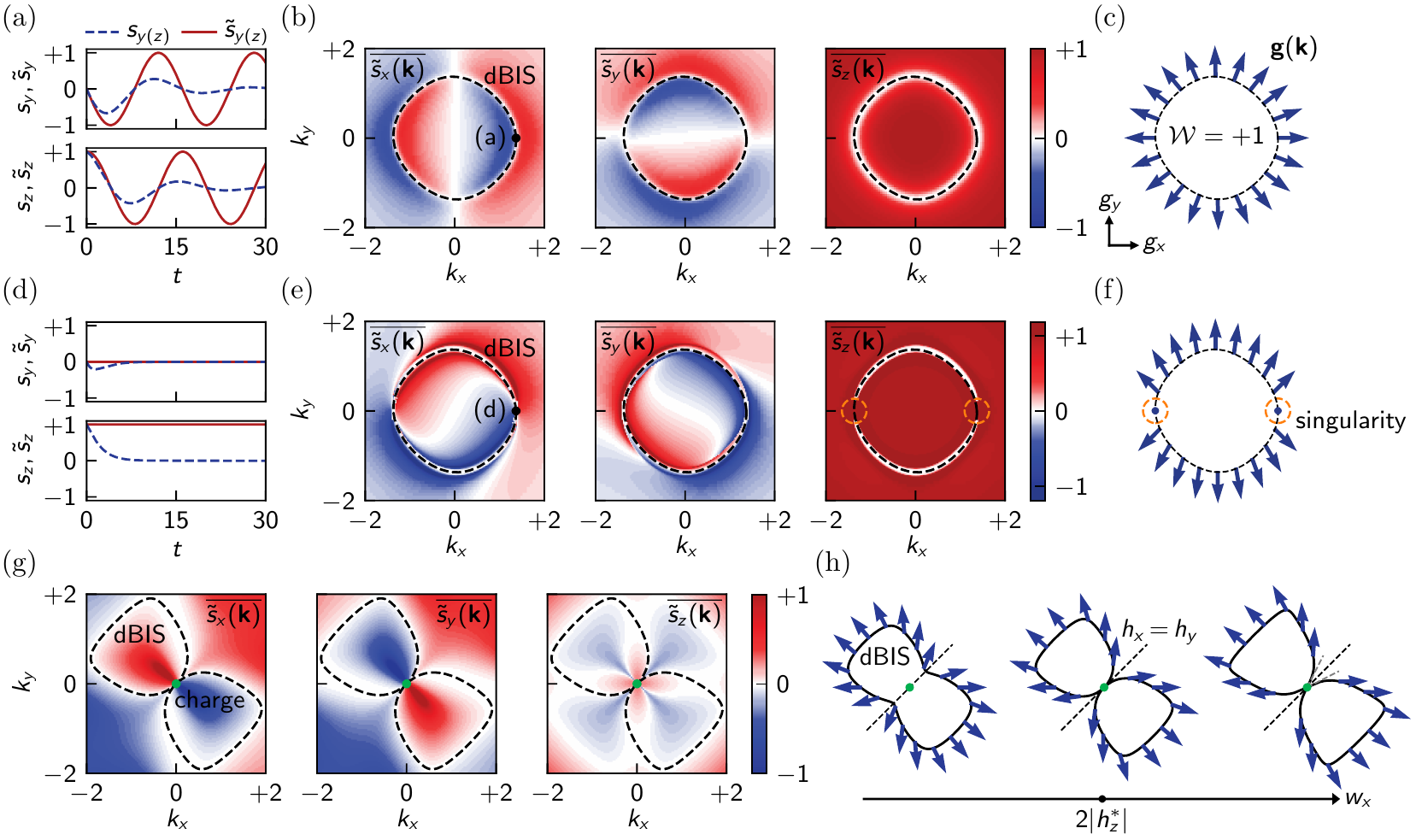}
\caption{Stability of the emergent dynamical topology.
(a) The rescaled dynamical spin polarization $\tilde{\boldsymbol{s}}(t)$
(red lines) and spin polarization $\boldsymbol{s}(t)$
(dashed blue lines) on dBIS momentum with $k_y=0$ marked by the black dot in (b).
The corresponding components $s_x,\tilde{s}_x$ are always zero.
(b) The time-averaged rescaled spin polarizations $\overline{\tilde{\boldsymbol{s}}(\mathbf{k})}$
under weak noise strength $w_{x}=0.05t_{0}$, $w_{y}=0$, $w_{z}=0.01t_{0}$.
The vanishing time averages $\overline{\tilde{\boldsymbol{s}}(\mathbf{k})}=0$ characterize the dBIS (dashed line).
(c) The dynamical pattern $\mathbf{g}(\mathbf{k})$ (blue arrows) exhibits nonzero topological number $\mathcal{W}=+1$.
(d) The rescaled dynamical spin polarization and spin polarization on dBIS at the momentum marked in (e).
(e) The time-averaged spin texture under strong noise strength $w_{x}=0.1t_{0}$, $w_{y}=0.05t_{0}$, $w_{z}=0.45t_{0}$. Singularities emerge
on dBIS momenta with nonzero time averaging $\overline{\tilde{\boldsymbol{s}}(\mathbf{k})}\neq 0$.
The spin dynamics at the marked momenta purely decays without oscillation.
(f) The dynamical pattern $\mathbf{g}(\mathbf{k})$ has singularities. Here we set $m_z=1.2t_0$ and $t_{\mathrm{so}}=0.2t_0$ for (a-f).
(g) The results for noise strength $w_x=1.6t_0$, $w_y=0$, $w_z=0.8t_0$, where the deformed dBIS connects to the topological charge (green dot) at $\mathbf{k}=0$. The dynamical topology also breaks down. Here we set $m_z=1.2t_0$ and $t_{\mathrm{so}}=2t_0$.
(h) The variation of dynamical pattern with $w_x$ increasing from zero while keeping $w_y=0$ and $w_z=0.8t_0$. When $w_x$ increases to $2|h^*_z|=1.6t_0$, the dBIS is deformed to the charge position where spin oscillation vanishes and the emergent topology breaks down. As $w_x$ further increases, more dBIS momenta are deformed to charge momentum.
\label{fig:figure2}}
\end{figure*}

The emergent topology characterizes the global feature of the dynamical spin texture which is normalized~\cite{Zhang2018}. For this we neglect the amplitude decay (quantified by $\mathrm{Re}\,\lambda_{\alpha}$) of the spin polarization to characterize the dynamical topology, except for the emergence of singularities, leading to the rescaled dynamical spin polarization
\begin{align}
\tilde{\boldsymbol{s}}(\mathbf{k},t) & \equiv\boldsymbol{s}_{0}(\mathbf{k})+\boldsymbol{s}_{+}(\mathbf{k})e^{-\mathrm{i}\omega(\mathbf{k})t} +\boldsymbol{s}_{-}(\mathbf{k})e^{+\mathrm{i}\omega(\mathbf{k})t}
\end{align}
This characterization requires the frequency $\omega(\mathbf{k})$ to be real.
Otherwise, spin dynamics is characterized by purely decay, as dominated by the noise, and the information of the topological phase of
post-quench Hamiltonian is scrambled for the indistinguishable decay modes with real $\lambda_\alpha$'s.
The emergent topology of the quench dynamics is characterized by the topological invariant $\mathcal{W}\equiv\frac{1}{2\pi}\oint_{{\rm dBIS}}\mathbf{g}(\mathbf{k})\mathrm{d}\mathbf{g}(\mathbf{k})$. Here $\mathbf{g}(\mathbf{k})$ denotes the directional
derivative of dynamical spin polarization in $x$-$y$ plane $\mathbf{g}(\mathbf{k})\equiv(1/\mathcal{N_{\mathbf{k}}})\partial_{k_{\perp}}(\overline{\tilde{s}_{x}(\mathbf{k})},\overline{\tilde{s}_{y}(\mathbf{k})})\vert_{{\rm dBIS}}$ normalized by $\mathcal{N}_{\mathbf{k}}$, with $k_{\perp}$ perpendicular to the dBISs and points to the side with $\boldsymbol{s}^L_0(\mathbf{k})\cdot\boldsymbol{s}(\mathbf{k},0)>0$. 
The invariant $\mathcal{W}$ represents the winding number of $\mathbf{g}(\mathbf{k})$ over dBISs. This invariant naturally reduces to the one for unitary quench dynamics if the noise is absent, which characterizes the topological phase of post-quench Hamiltonian~\cite{Zhang2018, Zhang2019b}.

\emph{Stability of the dynamical topology}.---Now we study the stability of dynamical topology, and consider first the weak noise regime, with the spin dynamics on dBIS exhibiting damped oscillation.
Accordingly $\tilde{\boldsymbol{s}}(\mathbf{k},t)$ on the dBISs oscillate around zero 
[see Fig.~\ref{fig:figure2}(a)], leading to vanishing time averaged spin polarization $\vert\overline{\tilde{\boldsymbol{s}}(\mathbf{k})}|_{\bold k\in\rm dBISs}=0$, see Fig.~\ref{fig:figure2}(b).
Near the dBISs, for the real and nonzero oscillation frequency $\omega(\mathbf{k})$ we have $\overline{\tilde{\boldsymbol{s}}(\mathbf{k})}=\boldsymbol{s}_{0}(\mathbf{k})$.
It follows that the dynamical field is obtained by
$\mathbf{g}(\mathbf{k})=\frac{1}{\mathcal{N}_{\mathbf{K}}}(s_{0,x}^{R}(\mathbf{k}),s_{0,y}^{R}(\mathbf{k}))\vert_{{\rm dBIS}}$
and explicitly reads~\cite{SM}
\begin{equation}
\mathbf{g}(\mathbf{k})=\frac{1}{\mathcal{N}_{\mathbf{k}}}(h_{x}(\mathbf{k})[1-\eta(\mathbf{k})],h_{y}(\mathbf{k})[1+\eta(\mathbf{k})])\vert_{{\rm dBIS}},
\end{equation}
where the deformation factor is $\eta=[(w_{y}-w_{x})/\mathbf{h}_{{\rm so}}^{2}]\times(w_{x}h_{x}^{2}/\mathbf{h}_{{\rm so}}^{2}+w_{y}h_{y}^{2}/\mathbf{h}_{{\rm so}}^{2}-w_{z})$.
The factor $\eta$ quantifies the deformation of $\mathbf{g}(\bold k)$ from the SO field $\mathbf{h}_{\mathrm{so}}$. It is seen that the noise cannot fully flip $\mathbf{g}$ with respect to $\mathbf{h}_{\mathrm{so}}$~\cite{SM}, which is crucial for the robustness of dynamical topology in the relatively weak noise regime. Finally, for the case with $w_x=w_y$ or without noise, the SO field is recovered.
These results are valid in general for the damped oscillating spin dynamics on the dBISs.

The stability of dynamical topology under weak noise can be examined by increasing the noise strength from zero.
Without noise, the invariant $\mathcal{W}$ reduces to the winding of SO field $\mathbf{h}_{\mathrm{so}}$ on the BISs,
which is clearly nontrivial.
As the noise increases to weak regime, no topological charge with $\mathbf{h}_{{\rm so}}(\mathbf{k})=0$ passes through the deformed dBISs~\cite{SM}.
Thus the winding of the dynamical field $\mathbf{g}(\mathbf{k})$ on the dBISs remains invariant. In Supplemental Material~\cite{SM},
we prove that while the $\mathbf{g}(\mathbf{k})$-field on the dBISs is locally deformed by the noise, its global topology, quantified by $\mathcal{W}$ is unchanged and still equivalent to
that of the SO field $\mathbf{h}_{\rm so}$, as long as dissipative spin
dynamics has finite minimal oscillation frequency over the dBISs, which mimics a bulk gap protecting the emergent dynamical phase against relatively weak noise.
In Fig.~\ref{fig:figure2}(c), we show the dynamical field $\mathbf{g}$ of QAH model under noise strengths $w_{x}=0.05t_{0}$, $w_{y}=0$, and $w_{z}=0.01t_{0}$. The quench dynamics exhibit nonzero topological number $\mathcal{W}=+1$.

The quench dynamics are qualitatively different in the strong noise regime. Two types of dynamical transitions are obtained, with numerical results shown in Fig.~\ref{fig:figure2}(d-h). For type-I, the spin dynamics at some momenta of dBIS purely decays without oscillation [Fig.~\ref{fig:figure2}(d)], even though the deformation of dBIS is relatively small. At these momenta, the eigenvalues $\lambda_\alpha$ of $\mathcal{L}(\mathbf{k})$ are all real, i.e. $\omega(\mathbf{k})$ is purely imaginary. The rescaled spin polarization $\tilde{\boldsymbol{s}}(\mathbf{k},t)=\sum_{\alpha=0,\pm}\boldsymbol{s}_{\alpha}(\mathbf{k})=\boldsymbol{s}(\mathbf{k},0)$ is trivial.
Thus the dBIS breaks down and the dynamical field $\mathbf{g}(\mathbf{k})$ can not be defined at these singular momenta in this regime [Fig.~\ref{fig:figure2}(e,f)]. For type-II, the noise strongly deforms the dBIS to connect to topological charges. At the crossing of charges with dBISs the dynamics vanish, leading to breakdown of the dynamical topology [Fig.~\ref{fig:figure2}(g,h)].
Thus the breakdown of the dynamical topology is associated with the emergence of singularities.

\begin{figure}[ht]
\includegraphics{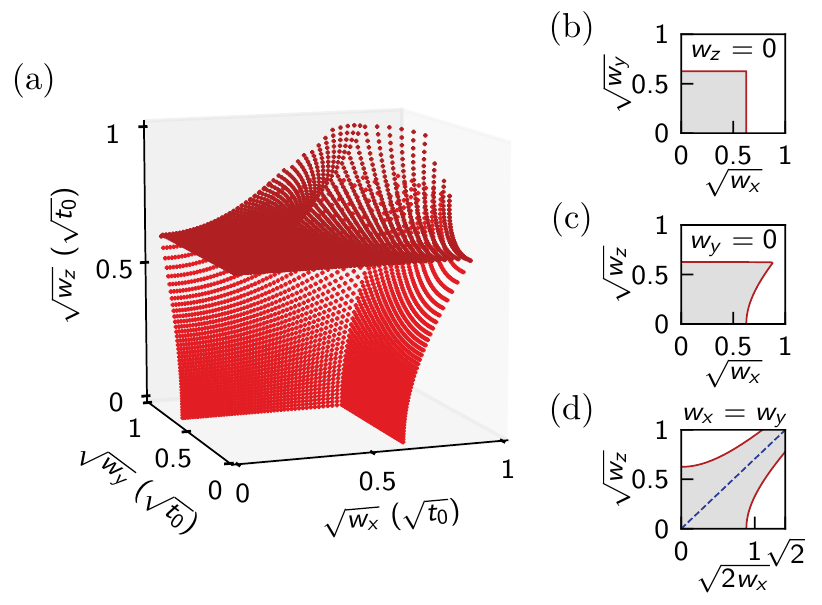}

\caption{Critical noise strength for QAH model. (a) The critical noise strength (red points), beyond which the
emergent topology breaks down. We show the slices with
$w_{z}=0$ and $w_{y}=0$ in (b) and (c) respectively. The slice with
$w_{x}=0$ is the same as (c). In the shadow region, the dynamical
topology is robust. (d) The slice with $w_{x}=w_{y}$.
The blue dashed line represents the sweet spot regime with
$w_{x}=w_{y}=w_{z}$. Here the parameters $m_{z}$
and $t_{{\rm so}}$ are the same as in Fig.~\ref{fig:figure1}. \label{fig:figure3}}
\end{figure}

\emph{Critical noise strength}.---The qualitative difference between the weak and strong noise regimes indicates that dynamical transitions occur at certain critical noise strengths. The dynamical topology is robust as long as the oscillation frequency of the dissipative spin dynamics is finite everywhere on dBISs, namely, the effective bulk gap of the emergent dynamical phase is not closed by the noise. Similar to the equilibrium topological phase, the dynamical transition here occurs when the minimal frequency vanishes. For type-II transition, the critical noise strength satisfies $|w_{y,c}-w_{x,c}|=2|h^{*}_{z}|$,
with $|h^{*}_z|$ the minimal value of $h_z$ at the topological charges where $\mathbf{h}_{\rm so}=0$. For $|w_{y}-w_{x}|\geq 2|h^{*}_{z}|$, the deformed dBISs connect to topological charges [Fig.~\ref{fig:figure2}(g,h)], where oscillations always vanish and the emergent topology breaks down~\cite{SM}. For the type-I transition, the critical noise strength $\boldsymbol{w}_{\rm c}=(w_{x,{\rm c}},w_{y,{\rm c}},w_{z,{\rm c}})$ is determined by
\begin{eqnarray}\label{critical}
\min_{\mathbf{k}\in\mathrm{dBISs}}\omega(\mathbf{k};\boldsymbol{w}_{\rm c})=0,
\end{eqnarray}
with $\omega(\mathbf{k})=2\sqrt{\mathbf{h}_{{\rm so}}^{2}(\mathbf{k})-[\lambda_{0}(\mathbf{k})/4-w_{z}]^{2}}$ and
$\lambda_{0}=2w_{x}h_{y}^{2}/\mathbf{h}_{{\rm so}}^{2}+2w_{y}h_{x}^{2}/\mathbf{h}_{{\rm so}}^{2}+2w_{z}$~\cite{SM}. Beyond critical value
the dynamical pattern becomes singular. Which type of transitions occurs first is parameter dependent.

Interestingly, from Eq.~\eqref{critical} a \emph{sweet spot} regime is predicted at $w_x=w_y=w_z$, in which case the dynamical topology survives at arbitrarily strong noise. In this case the dBISs reduced to the original BISs and the spin oscillations are fully determined by the SO field on BISs, even the noise is present. The noise only induces a decay factor for the spin evolution but the emergent dynamical topology is not affected. This is an exotic feature of the dynamical topology, not similar to equilibrium counterparts.
The critical noise strength for QAH model is shown in Fig.~\ref{fig:figure3}(a), with the slices $w_{z}=0$, $w_{y}=0$ and $w_x=w_y$
in Fig.~\ref{fig:figure3}(b-d) respectively. The topological region is bounded by
these critical points, in which the dynamical topology is robust against noise.
As the noise strength increases, the topological region shrinks and approaches the sweet spot regime [see Fig.~\ref{fig:figure3}(d)].

\emph{Conclusion and outlook}.---
We have studied the effect of dynamical noise on the quench-induced emergent dynamical topology, with novel results being predicted. We showed that the dynamical topology is robust against weak dynamical noise, and protected by an effective bulk gap, the minimal oscillation frequency of dynamics over dBIS. Two types of dynamical transitions are predicted under strong dynamical noise, with critical points being obtained. Interestingly, we predicted a novel {\em sweet spot} regime, 
in which case the dynamical topology survives in arbitrarily strong noise regime. Motivated by this study the interesting open questions can be further asked. In particular, as in equilibrium topological phases the nontrivial physics have been extensively studied, e.g. the emergence of gapless boundary modes, topological defect modes, topological responses under external field, etc., can the similar nontrivial topological physics be obtained in the emergent dynamical topology? These issues are novel and deserve future studies.

This work was supported by National Natural Science Foundation
of China (Grants No. 11825401, No. 11761161003, and
No. 11921005), the National Key R\&D Program of China
(Project No. 2016YFA0301604), and the Strategic Priority Research Program of
Chinese Academy of Science (Grant No. XDB28000000).




\renewcommand{\theequation}{S\arabic{equation}}
\setcounter{equation}{0}  
\renewcommand{\thefigure}{S\arabic{figure}}
\setcounter{figure}{0}  

\onecolumngrid
\flushbottom
\newpage

\begin{center}\large
\textbf{Supplemental Material}
\end{center}
In this Supplemental Material, we provide more details on the Liouvillian superoperator
and the spin dynamics on the dynamical band-inversion surfaces (dBISs).
Then we prove the robustness of the emergent dynamical topology under weak noise.
Finally, we derive the critical conditions for noise strength.

\section{The Liouvillian superoperator}

For the Lindblad master equation in main text, the Liouvillian superoperator $\mathcal{L}(\mathbf{k})$ is defined as
\begin{equation}
\frac{\mathrm{d}}{\mathrm{d}t}\rho(\mathbf{k},t)=\mathcal{L}(\mathbf{k})[\rho(\mathbf{k},t)],\qquad \mathcal{L}(\mathbf{k})[\rho(\mathbf{k},t)]\equiv-\mathrm{i}[H(\mathbf{k}),\rho(\mathbf{k},t)]+\sum_{i=x,y,z}w_{i}[\sigma_{i}\rho(\mathbf{k},t)\sigma_{i}-\rho(\mathbf{k},t)].
\end{equation}
For the two-band post-quench Hamiltonian $H(\mathbf{k})=\mathbf{h}(\mathbf{k})\cdot\boldsymbol{\sigma}$,
the density matrix generally takes the form $\rho(\mathbf{k},t)=\frac{1}{2}[I+\boldsymbol{n}(\mathbf{k},t)\cdot\boldsymbol{\sigma}]$.
Therefore, we have $\frac{1}{2}\frac{\mathrm{d}}{\mathrm{d}t}\boldsymbol{n}(\mathbf{k},t)\cdot\boldsymbol{\sigma}=[\mathbf{h}(\mathbf{k})\times\boldsymbol{n}(\mathbf{k},t)]\cdot\boldsymbol{\sigma}-\sum_{i,j=x,y,z;j\neq i}w_{i}n_{j}(\mathbf{k},t)\sigma_{j}$.
In the basis $(\sigma_x,\sigma_y,\sigma_z)$, the Liouvillian superoperator is given by
\begin{equation}
\mathcal{L}(\mathbf{k})=\begin{bmatrix}-2(w_{y}+w_{z}) & -2h_{z}(\mathbf{k}) & 2h_{y}(\mathbf{k})\\
2h_{z}(\mathbf{k}) & -2(w_{x}+w_{z}) & -2h_{x}(\mathbf{k})\\
-2h_{y}(\mathbf{k}) & 2h_{x}(\mathbf{k}) & -2(w_{x}+w_{y})
\end{bmatrix}.\label{eq:Liouvillian superoperator_S}
\end{equation}

\section{Spin dynamics on the dynamical band-inversion surfaces}

With the Liouvillian superoperator, the equation of motion for the stochastic averaged spin polarization
$\boldsymbol{s}(\mathbf{k},t)=\mathrm{Tr}[\rho(\mathbf{k},t)\boldsymbol{\sigma}]$ reads
\begin{equation}
\dot{\boldsymbol{s}}(\mathbf{k},t)=\mathcal{L}(\mathbf{k})\boldsymbol{s}(\mathbf{k},t).
\end{equation}
Then the stochastic averaged spin polarization is given by $\boldsymbol{s}(\mathbf{k},t)=e^{\mathcal{L}(\mathbf{k})t}\boldsymbol{s}(\mathbf{k},0)$ and generally takes the form
\begin{equation}
\boldsymbol{s}(\mathbf{k},t)=\boldsymbol{s}_{0}(\mathbf{k})e^{-\lambda_{0}(\mathbf{k})t}+\boldsymbol{s}_{+}(\mathbf{k})e^{-[\lambda_{1}(\mathbf{k})+\mathrm{i}\omega(\mathbf{k})]t}+\boldsymbol{s}_{-}(\mathbf{k})e^{-[\lambda_{1}(\mathbf{k})-\mathrm{i}\omega(\mathbf{k})]t},
\end{equation}
with $\boldsymbol{s}_{\alpha}(\mathbf{k})=[\boldsymbol{s}_{\alpha}^{L}(\mathbf{k})\cdot\boldsymbol{s}(\mathbf{k},0)]\boldsymbol{s}_{\alpha}^{R}(\mathbf{k})$
for $\alpha=0,\pm$. Here $\boldsymbol{s}_{\alpha}^{L(R)}$
are the left (right) eigenvectors for the Liouvillian superoperator $\mathcal{L}(\mathbf{k})$, namely
$\mathcal{L}^{T}\boldsymbol{s}^{L}_{\alpha}=-\lambda_{\alpha}\boldsymbol{s}^{L}_{\alpha}$
and $\mathcal{L}\boldsymbol{s}^{R}_{\alpha}=-\lambda_{\alpha}\boldsymbol{s}^{R}_{\alpha}$.
The corresponding eigenvalues are determined by $\lambda_{0}$ and $\lambda_{\pm}=\lambda_{1}\pm\mathrm{i}\omega$.
Note that $\boldsymbol{s}^{L(R)}_0$ is real.
These eigenvectors satisfy $\boldsymbol{s}_{m}^{L}(\mathbf{k})\cdot\boldsymbol{s}_{n}^{R}(\mathbf{k})=\delta_{mn}$.
Here we focus on the spin dynamics on dBISs defined as $\mathrm{dBIS}\equiv\{\mathbf{k}\vert\boldsymbol{s}_{0}^{L}(\mathbf{k})\cdot\boldsymbol{s}(\mathbf{k},0)=0\}$.

\subsection{Evolution within the plane perpendicular to the spin-orbit axis}

To prove that the spin polarization on dBISs evolves within the plane perpendicular to the spin-orbit (SO) axis $\mathbf{h}_{\rm so}\equiv(h_x,h_y)$,
we notice that the spin component $s_{\parallel}(\mathbf{k},t)=\boldsymbol{s}^{L}_{0}(\mathbf{k})\cdot\boldsymbol{s}(\mathbf{k},t)$
satisfies
\begin{equation}
\dot{s}_{\parallel}(\mathbf{k},t)=-\lambda_{0}(\mathbf{k})s_{\parallel}(\mathbf{k},t).
\end{equation}
On the dBISs with $\boldsymbol{s}_{0}^{L}(\mathbf{k})\cdot\boldsymbol{s}(\mathbf{k},0)=0$,
it is obvious that the component $s_{\parallel}$ always vanishes, namely $s_{\parallel}(\mathbf{k},t)=0$.
Thus the spin polarization evolves within the plane perpendicular to the component $s_{\parallel}$.
On the other hand, for our initial state $\ket{\psi(\mathbf{k},t=0)}=\ket{\uparrow}$, i.e. $\boldsymbol{s}(\mathbf{k},0)=(0,0,1)$,
$\boldsymbol{s}^L_0$ on dBISs generally takes the form $\boldsymbol{s}^L_0=(\alpha,\beta,0)$ satisfying
\begin{equation}\label{eq:sL0}
(\alpha,\beta)\begin{pmatrix}-w_{y}+w_x{} & -2h_{z} & 2h_{y}\\
2h_{z} & w_{y}-w_{x} & -2h_{x}
\end{pmatrix}=-(\lambda_{0}-w_x-w_y-2w_z)(\alpha,\beta,0).
\end{equation}
Thus we have $\boldsymbol{s}^L_0\sim(h_x,h_y,0)$ on the dBISs and
\begin{equation}
s_{\parallel}=h_xs_x+h_ys_y,
\end{equation}
which is parallel to the SO axis $\mathbf{h}_{\rm so}$ in the Bloch sphere.
Therefore, on the dBISs the spin polarization evolves
within the plane perpendicular to the SO axis $\mathbf{h}_{\rm so}$, which is valid both for weak noise and strong noise.

\subsection{Explicit expression for dBISs}

With $\boldsymbol{s}^L_0\sim(h_x,h_y,0)$ on the dBISs, we can solve the explicit expression for dBISs, which is given by
\begin{equation}\label{eq:dBIS_S}
h_z(\mathbf{k})=(w_y-w_x)\frac{h_x(\mathbf{k})h_y(\mathbf{k})}{h^2_x(\mathbf{k})+h^2_y(\mathbf{k})}.
\end{equation}
Here the dBISs are independent of the noise $w_z$ but dependes on the difference between $w_y$ and $w_x$, as seen from the left hand side of Eq.~(\ref{eq:sL0}).
Besides, the eigenvalue $\lambda_0$ reads
\begin{equation}
\lambda_0(\mathbf{k})=2w_x\frac{h^2_y(\mathbf{k})}{h^2_x(\mathbf{k})+h^2_y(\mathbf{k})}+2w_y\frac{h^2_x(\mathbf{k})}{h^2_x(\mathbf{k})+h^2_y(\mathbf{k})}+2w_z.
\end{equation}

\section{Robustness of the emergent dynamical topology}
The emergent dynamical topology is defined as
\begin{equation}
\mathcal{W}\equiv\frac{1}{2\pi}\oint_{\rm dBIS}\mathbf{g}(\mathbf{k})\mathrm{d}\mathbf{g}(\mathbf{k}),
\end{equation}
where the dynamical field $\mathbf{g}(\mathbf{k})$ is given by
$\mathbf{g}(\mathbf{k})\equiv\frac{1}{\mathcal{N_{\mathbf{k}}}}\partial_{k_{\perp}}(\overline{\tilde{s}_{x}(\mathbf{k})},\overline{\tilde{s}_{y}(\mathbf{k})})\vert_{{\rm dBIS}}$,
and $\overline{\tilde{\boldsymbol{s}}(\mathbf{k})}$ is the time-averaged dynamical spin polarization (see main text).
Here $k_\perp$ is perpendicular to the dBISs and $\mathcal{N}_{\mathbf{k}}$ is a normalization factor.
The dynamical invariant $\mathcal{W}$ represents the winding of $\mathbf{g}(\mathbf{k})$ over dBISs.
Here we prove the robustness of the dynamical topology under weak noise.

\subsection{The dynamical field on dBISs}

For the weak noise with damped oscillating spin dynamics on the dBISs, the time-averaged dynamical spin polarization is given by
$\overline{\tilde{\boldsymbol{s}}(\mathbf{k})}=\boldsymbol{s}_0(\mathbf{k})=[\boldsymbol{s}^L_0(\mathbf{k})\cdot\boldsymbol{s}(\mathbf{k},0)]\boldsymbol{s}^R_0(\mathbf{k})$
and the dBISs are characterized by the momenta with $\overline{\tilde{\boldsymbol{s}}(\mathbf{k})}=0$.
We have
\begin{equation}
\partial_{k_\perp}\overline{\tilde{s}_\alpha(\mathbf{k})}\vert_{\rm dBIS}=(\partial_{k_\perp}[\boldsymbol{s}^L_0(\mathbf{k})\cdot\boldsymbol{s}(\mathbf{k},0)])s^R_{0,\alpha}(\mathbf{k})\vert_{\rm dBIS}
\end{equation}
for $\alpha=x,y$ due to $\boldsymbol{s}^L_0(\mathbf{k})\cdot\boldsymbol{s}(\mathbf{k},0)=0$ on the dBISs.
For topologically nontrivial post-quench Hamiltonians $H(\mathbf{k})$ with band-inversion surfaces (BISs)~\cite{Zhang2018S},
$\boldsymbol{s}^L_0(\mathbf{k})\cdot\boldsymbol{s}(\mathbf{k},0)$ generally has opposite signs in the two sides of dBISs,
leading to nonzero $\partial_{k_\perp}[\boldsymbol{s}^L_0(\mathbf{k})\cdot\boldsymbol{s}(\mathbf{k},0)]\vert_{\rm dBIS}$.
Here we choose $k_\perp$ pointing to the side of dBISs with $\boldsymbol{s}^L_0(\mathbf{k})\cdot\boldsymbol{s}(\mathbf{k},0)>0$.
After normalization, the dynamical field is determined by the right eigenvector $\boldsymbol{s}^R_0$ of the Liouvillian superoperator,
\begin{equation}
\mathbf{g}(\mathbf{k})=\frac{1}{\mathcal{N}_{\bf k}}(s^R_{0,x}(\mathbf{k}),s^R_{0,y}(\mathbf{k}))\vert_{\rm dBIS},
\end{equation}
where the $x,y$ components of the right eigenvector $\boldsymbol{s}^R_0(\mathbf{k})$ on the dBISs are given by
\begin{equation}
(s^R_{0,x},s^R_{0,y})\sim(h_x(1-\eta),h_y(1+\eta))\vert_{\rm dBIS}
\end{equation}
with the deformation factor $\eta=\frac{w_y-w_x}{h^2_x+h^2_y}\left(w_x\frac{h^2_x}{h^2_x+h^2_y}+w_y\frac{h^2_y}{h^2_x+h^2_y}-w_z\right)$.
Thus the dynamical field $\mathbf{g}$ is generally deformed from the SO field $\mathbf{h}_{\mathrm{so}}$ on the dBISs.
Particularly, we have $\mathbf{g}\sim(0,h_y)$ for the momenta with $h_x=h_z=0$ but $h_y\neq 0$
and $\mathbf{g}\sim(h_x,0)$ for the momenta with $h_y=h_z=0$ but $h_x\neq 0$
[with the condition $h^2_x+h^2_y>(\lambda_0/4-w_z)^2$ for weak noise, it is easy to find $1\pm\eta>0$ for these two cases respectively].
It is obvious that the dynamical field $\mathbf{g}(\mathbf{k})$ on the dBISs is nonzero, since $1-\eta$ and $1+\eta$ cannot be zero simultaneously.
Furthermore, the noise cannot fully flip the SO field, namely $\mathbf{g}\not\propto-\mathbf{h}_{\mathrm{so}}$, otherwise we must have $1-\eta=1+\eta<0$, which is impossible.
These results are valid for arbitrary noise strength with damped oscillating spin dynamics on the dBISs.

\subsection{The winding of SO field on the BISs and dBISs}

Without noise, the invariant $\mathcal{W}$ reduces to the winding of the SO field $\mathbf{h}_{\rm so}$ on the band-inversion surfaces defined as
$\mathrm{BIS}\equiv\{\mathbf{k}\vert h_z(\mathbf{k})=0\}$, which is clearly nontrivial and counts the topological charges with
$\mathbf{h}_{\rm so}(\mathbf{k})=0$ enclosed by the BISs~\cite{Zhang2018S,Zhang2019S}. Here we show that as the noise increases from zero to weak regime,
no topological charge passes through the deformed dBISs.
The proof is given as follows. For the momenta of topological charges, the spin dynamics with initial state
$\boldsymbol{s}(t=0)=(0,0,1)$ is given by $s_x(t)=s_y(t)=0$ and $s_z(t)=e^{-2(w_x+w_y)t}$,
and the corresponding dynamical spin polarization reads $\tilde{\boldsymbol{s}}(t)=(0,0,1)$.
If we gradually increase the strength of noise from zero to some weak noise,
the dBISs shall be deformed from the BISs to the final ones but no topological charge passes through
them, otherwise there will be momenta on the dBISs with
$\overline{\tilde{\boldsymbol{s}}(\mathbf{k})}\neq 0$ for the intermediate weak noise, which is impossible.
Thus no topological charge passes through the deformed dBISs.

Consequently, the topological
charges enclosed by the dBISs are the same as those enclosed by the BISs,
and the winding of the SO field $\mathbf{h}_{\rm so}$ on the dBISs remains invariant in the weak noise regime.

\subsection{Topological equivalence between the dynamical field and SO field on dBISs}

We now prove the robustness of the dynamical topology by showing that the winding of the dynamical field $\mathbf{g}$ on the dBISs is equivalent to that of the SO field $\mathbf{h}_{\rm so}$
on the dBISs. We can introduce two auxiliary gapped Hamiltonians defined on the one-dimensional dBISs,
\begin{equation}
H_1(\mathbf{k})=g_x(\mathbf{k})\sigma_x+g_y(\mathbf{k})\sigma_y,\qquad H_2(\mathbf{k})=h_x(\mathbf{k})\sigma_x+h_y(\mathbf{k})\sigma_y
\end{equation}
with $\mathbf{k}\in\mathrm{dBIS}$,
of which the winding number characterizes the winding of the dynamical field $\mathbf{g}$ and the SO
field $\mathbf{h}_{\rm so}$ respectively. Here we can choose $(g_x,g_y)$ as $(h_x(1-\eta),h_y(1+\eta))$ without normalization, which does not
change the winding of the dynamical field.
Consider the continuous deformation
$H(\mathbf{k},q)=qH_1(\mathbf{k})+(1-q)H_2(\mathbf{k})$ with $0\leq q \leq 1$, we have
\begin{align}
[qH_{1}+(1-q)H_{2}]^{2} & =[qh_{x}(1-\eta)+(1-q)h_{x}]^{2}+[qh_{y}(1+\eta)+(1-q)h_{y}]^{2}\nonumber \\
 & =(1-q\eta)^{2}h_{x}^{2}+(1+q\eta)^{2}h_{y}^{2}.
\end{align}
Since $1-q\eta$ and $1+q\eta$ cannot be zero simultaneously, we must have $[qH_{1}+(1-q)H_{2}]^{2}>0$
for $h_x\neq 0$ and $h_y\neq 0$. For the case with $h_x=0$ or $h_y=0$, as discussed above it is obvious that
$[qH_{1}+(1-q)H_{2}]^{2}>0$. Therefore, $H(\mathbf{k},q)$ is gapped for $0\leq q \leq 1$, indicating
that $H_1$ and $H_2$ are topologically equivalent.

This proves that for the weak noise with damped oscillation for spin dynamics on dBISs,
the winding of the corresponding dynamical field $\mathbf{g}(\mathbf{k})$
is topologically equivalent to that of the SO field and remains invariant,
demonstrating the robustness of the emergent dynamical topology.

\section{Critical noise strength}

As shown in the main text, there exist two types of dynamical transitions for the emergent dynamical topology.
In this section, we derive the corresponding critical noise strength.

\subsection{Oscillation frequency on the dBISs and type-I critical noise strength}

For type-I dynamical transition, the deformation of dBISs is relatively small,
but the minimal oscillation frequency over the dBISs vanishes for noise beyond the critical noise strenth.
The critical noise strength can be determined from the oscillation frequency on dBISs by solving the equation $\det(\mathcal{L}+\lambda I)=0$, namely $\lambda^3+b\lambda^2+c\lambda+d=(\lambda-\lambda_0)[\lambda^2+(b+\lambda_0)\lambda+\lambda^2_0+b\lambda_0+c]=0$ with $b=-4(w_x+w_y+w_z)$, $c=4[(w_x+w_z)(w_x+w_y)+(w_y+w_z)(w_x+w_y)+(w_x+w_z)(w_y+w_z)+h^2_x+h^2_y+h^2_z]$ and $d=-8[h^2_x(w_y+w_z)+h^2_y(w_x+w_z)+h^2_z(w_x+w_y)+(w_x+w_y)(w_x+w_z)(w_y+w_z)]$. We obtain $\lambda^2_\pm+(b+\lambda_0)\lambda_\pm+\lambda^2_0+b\lambda_0+c$, and the oscillation frequency on dBISs is given by
\begin{equation}
\omega(\mathbf{k}) = 2\sqrt{h^2_x(\mathbf{k})+h^2_y(\mathbf{k})-[\lambda_0(\mathbf{k})/4-w_z]^2}.
\end{equation}
For strong noise, we may have $h^2_x+h^2_y\leq [\lambda_0/4-w_z]^2$ and $\omega$ is purely imaginary for some dBIS momenta,
namely the minimal frequency over dBISs vanishes. In this case, the spin dynamics on such dBIS momenta purely decay without oscillation, and the emergent dynamical topology breaks down.
Thus the critical noise strength $\boldsymbol{w}_{\rm c}=(w_{x,\mathrm{c}}, w_{y,\mathrm{c}}, w_{z,\mathrm{c}})$
for type-I dynamical transition is determined by
\begin{eqnarray}
\min_{\mathbf{k}\in\mathrm{dBISs}}\omega(\mathbf{k};\boldsymbol{w}_{\rm c})=0.
\end{eqnarray}

\subsection{dBISs connecting to the topological charges and type-II critical noise strength}

For type-II dynamical transition, the dBISs are deformed dramatically and even connect to the topological charges,
where the dynamical topology breaks down.
To see this, we consider a continuous path $P$ with fixed
$h_xh_y/(h^2_x+h^2_y)$ connecting the corresponding momentum on BISs
with $h_z=0$ and the topological charge with $\mathbf{h}=(0,0,h^*_z)$. This path always exists, as seen in the $h_x\text{-}h_y$ plane.
As $|w_y-w_x|$ increases, the dBIS momentum with increasing $|h_z|=\frac{|h_xh_y|}{h^2_x+h^2_y}|w_y-w_x|$ will
move along the path $P$. Due to $h^2_x+h^2_y\geq 2|h_xh_y|$, the dBIS momentum varying most rapidly belongs to the path $P$ with $|h_xh_y|/(h^2_x+h^2_y)=\frac{1}{2}$, namely $|h_x|=|h_y|$.
For this path $P$, we have $|h_z|=|w_y-w_x|/2$.
When $|w_y-w_x|$ increases to $2|h^*_z|$, we have $h_z=h^*_z$ on the dBIS momentum, indicating that the deformed dBISs connect to the topological charge.
Actually, this is indeed the case. We can examine the spin dynamics for the topological charge with $\mathbf{h}_{\rm so}=0$,
\begin{equation}
\mathcal{L}=\begin{bmatrix}-2(w_{y}+w_{z}) & -2h^*_{z} & 0 \\
2h^*_{z} & -2(w_{x}+w_{z}) & 0 \\
0 & 0 & -2(w_{x}+w_{y})
\end{bmatrix}.
\end{equation}
The oscillation frequency on the topological charge is nonzero for $|w_y-w_x|<2|h^*_z|$,
for which the topological charge momentum cannot belong to the dBISs due to
$\boldsymbol{s}^L_{0}\cdot\boldsymbol{s}(0)\neq 0$ for the real eigenvalue $\lambda_0$.
However, for $|w_y-w_x|\geq 2|h^*_z|$, the oscillation frequency on the charge vanishes and we have
the eigenvector $\boldsymbol{s}^{L}_0\sim(\alpha,\beta,0)$ for real $\lambda_0\neq 2(w_x+w_y)$,
which sastifies the dBIS condition $\boldsymbol{s}^L_{0}\cdot\boldsymbol{s}(0)= 0$.

Therefore, the dBISs connect to the topological charge for $|w_y-w_x|\geq 2|h^*_z|$. However, for this case
the oscillation frequency always vanishes and the dynamical topology breaks down.
This gives us the type-II critical noise strength
\begin{equation}
|w_{y,c}-w_{x,c}|=2|h^*_z|,
\end{equation}
with $|h^*_z|$ the minimum value of $|h_z|$ at topological charges.


\end{document}